\documentclass{JHEP3} 

\JHEPspecialurl{http://jhep.sissa.it/JOURNAL/JHEP3.tar.gz}

\usepackage{epsfig,multicol,bbm}

\newcommand\fverb{\setbox\fverbbox=\hbox\bgroup\verb}
\newcommand\fverbdo{\egroup\medskip\noindent%
                        \fbox{\unhbox\fverbbox}\ }
\newcommand\fverbit{\egroup\item[\fbox{\unhbox\fverbbox}]}
\newbox\fverbbox

\usepackage{graphicx}
\usepackage{amsmath}
\usepackage{amsfonts}
\usepackage{amssymb}
\def\rbt{R_{b/t}}
\def\what{\widehat}
\def\cabb{C_{ab\anti b}}
\def\catt{C_{at\anti t}}
\def\camumu{C_{a\mu^-\mu^+}}
\def\catautau{C_{a\tau^-\tau^+}}
\def\cmax{\cabb^{\rm max}}

\def\cta{\cos\theta_A}
\def\ctamax{\cta^{\rm max}}
\def\sta{\sin\theta_A}

\def\ma{m_a}
\def\mh{m_h}
\def\mhh{m_H}
\def\mhp{m_{h^+}}
\def\hi{h_1}
\def\hii{h_2}
\def\hiii{h_3}
\def\ai{a_1}
\def\aii{a_2}
\def\mhi{m_{h_1}}

\def\mai{m_{a_1}}
\def\maii{m_{\aii}}

\def\mtau{m_\tau}

\def\mb{m_b}
\def\mups{m_\Upsilon}
\def\beq{\begin{equation}}
\def\eeq{\end{equation}}
\def\bea{\begin{eqnarray}}
\def\eea{\end{eqnarray}}

\def\lsim{\mathrel{\raise.3ex\hbox{$<$\kern-.75em\lower1ex\hbox{$\sim$}}}}
\def\gsim{\mathrel{\raise.3ex\hbox{$>$\kern-.75em\lower1ex\hbox{$\sim$}}}}
\def\ifmath#1{\relax\ifmmode #1\else $#1$\fi}

\def\br{BR}
\def\gev{~{\mbox{GeV}}}

\def\to{\rightarrow}

\def\calo{{\cal O}}

\def\anti{\overline}

    \def\fillboxx#1#2{\hbox to #1{\vbox to #2{\vfil}\hfil}   }

\def\tauptaum{\tau^+\tau^-}

\def\gev{~{\rm GeV}}
\def\gam{\gamma}
\def\tanb{\tan\beta}
\def\cotb{\cot\beta}

\def\anti{\overline}

\def\epem{e^+e^-}
\def\etal{{\it et al.}}
\newcommand{ \slashchar }[1]{\setbox0=\hbox{$#1$}   
   \dimen0=\wd0                                     
   \setbox1=\hbox{/} \dimen1=\wd1                   
   \ifdim\dimen0>\dimen1                            
      \rlap{\hbox to \dimen0{\hfil/\hfil}}          
      #1                                            
   \else                                            
      \rlap{\hbox to \dimen1{\hfil$#1$\hfil}}       
      /                                             
   \fi}     

\title{A light CP-odd Higgs boson and the muon anomalous magnetic
  moment}

\author{John
  F. Gunion \\ Department of Physics, University of California, Davis, CA 95616}

\abstract{ We amalgamate the many experimental limits on the $ab\anti
  b$ coupling of a light CP-odd Higgs boson, $a$, including
  model-dependence coming from the ratio of the $at\anti t$ to the $a
  b\anti b$ coupling.  We then employ these limits to analyze the
  extent to which a light $a$ can make a significant contribution to
  the discrepancy, $\Delta a_\mu$, between the experimentally observed
  $a_\mu$ and that predicted by the standard model. In a
  ``model-independent'' framework and in the context of a general
  two-Higgs-doublet model this is a significant possibility. In
  contrast, the minimal supersymmetric model is too strongly
  constrained (after combining experimental and theoretical input) to
  allow a CP-odd-$a$ explanation of $\Delta a_\mu$.  The
  next-to-minimal supersymmetric model allows more freedom and the
  light $a$ of the model could explain the full $\Delta a_\mu$ if
  $9.2\gev<\ma<12\gev $, or contribute substantially for larger $\ma$,
  if $\tanb$ is large.  } 

\received{} 

\accepted{}
\begin{document}

There have been numerous studies~\cite{Chang:2000ii,Dedes:2001nx,Cheung:2001hz,Chen:2001kn,Wu:2001vq,Krawczyk:2001pe,Arhrib:2001xx,Krawczyk:2002df,Cheung:2003pw,Kong:2004um,Domingo:2007dx,Domingo:2008bb} of the extent to which the Higgs
sector could contribute to the anomalous magnetic moment of the muon,
$a_\mu$,
with current focus on whether it could be used
to explain some portion of the now $\sim 3\sigma$ positive deviation
of $a_\mu$ with respect to the Standard Model (SM) prediction. The
numerical deviation is variously quoted as $\Delta a_\mu\sim (27.5\pm
8.4)\times 10^{-10}$~\cite{Davier:2007ua} or $(27.7\pm 9.3)\times
10^{-10}$~\cite{Domingo:2008bb}.  It is becoming increasingly likely
that this deviation can only be explained by new physics of some kind
and a beyond-the-standard-model Higgs sector has always been a prime
candidate. 

Precision electroweak data and direct LEP limits on a light CP-even
scalar suggest that it should have SM-like couplings and substantial
mass, in which case its contribution to $a_\mu$ will only be of order
$few\times 10^{-11}$. Thus, we will focus on the possible
contribution, $\delta a_\mu$, of a light CP-odd Higgs boson, $a$, of a
CP-conserving Higgs sector, for which it is
critical~\cite{Chang:2000ii} to include the two-loop Barr-Zee type
diagrams~\cite{Barr:1990vd} since the one-loop $a_\mu$ contribution is
negative whereas the two-loop contribution is positive in popular
models.

Of particular interest is the $\ma<2\mb$ region, for which a light
Higgs, $h$, with SM-like $WW$, $ZZ$ and fermionic couplings can have 
mass $\mh\sim 100\gev$ while still being consistent with LEP data by
virtue of $h\to aa\to 4\tau$ decays being
dominant~\cite{Dermisek:2005ar,Dermisek:2005gg,Dermisek:2006wr,Dermisek:2007yt} (see
also \cite{Chang:2005ht,Chang:2008cw}).  Such a Higgs provides perfect
agreement with the rather compelling precision
electroweak constraints, and for $\br(h\to aa)\gsim 0.75$ also provides an
explanation for the $\sim 2.3\sigma$ excess observed at LEP in $\epem
\to Z b\anti b$ in the region $M_{b\anti b}\sim 100\gev$.  We term
this the ``ideal'' Higgs scenario. More generally, we will only consider
models for which the $ZZh$ coupling is SM-like (implying zero
$Zha$ coupling and therefore no lower limits on $\ma$ coming
from $\epem\to ha$ at LEP) and $\mh$ is such as to give good agreement
with precision electroweak data.

Possible contributions to $a_\mu$ by the $a$ depend crucially on the
$a\mu^-\mu^+$, $ab\anti b$ and $at\anti t$ couplings defined via
\beq
{\cal L}_{af\anti f}\equiv i C_{af\anti f}{ig_2m_f\over2m_W}\anti f \gamma_5 f
a\,.
\label{cabbdef}
\eeq 
We assume a Higgs model in which $\camumu=\catautau=\cabb$, as typified by a
two-Higgs-doublet model (2HDM) of either type-I or type-II (a 2HDM
contains Higgs bosons $h$, $H$ with $\mhh>\mh$, $a$ and $h^+$), or
more generally if the lepton and quark masses are generated by the
same combination of Higgs fields. (Much larger values of $a_\mu$
relative to those we find below are possible in models in which
$r=(\camumu=\catautau) /\cabb\gg 1$ --- such models include those in which the
muon and tau masses are generated by different Higgs fields than the $b$ mass.
For $r\neq 1$, our results for $\delta a_\mu$ should be rescaled by
$r$.)  In a 2HDM of type-II and in the MSSM, $\camumu=\cabb=\tanb$
(where $\tanb=h_u/h_d$ is the ratio of the vacuum expectation values
for the doublets giving mass to up-type quarks vs.  down-type quarks)
and $\catt=\cotb$.  In the NMSSM the expressions for $\camumu=\cabb$
and $\catt$ include an additional factor discussed later. In a type-I
2HDM, $\camumu=\catautau=\cabb=-\catt=-\cotb$. In the most general Higgs model,
$\camumu$, $\catautau$, $\cabb$ and $\catt$ will be more complicated functions of
the vevs of the Higgs fields and the structure of the Yukawa
couplings. In this paper, we assume $\camumu=\catautau=\cabb$ but allow for
general values of $\rbt^2\equiv \cabb/\catt$.  We consider only
positive values of $\rbt^2$ since only these are of relevance for explaining the
observed positive $\Delta a_\mu$ and positive values are typical of
most models.

\begin{figure}[h!]
\begin{center}
\includegraphics[width=0.65\textwidth,angle=90]{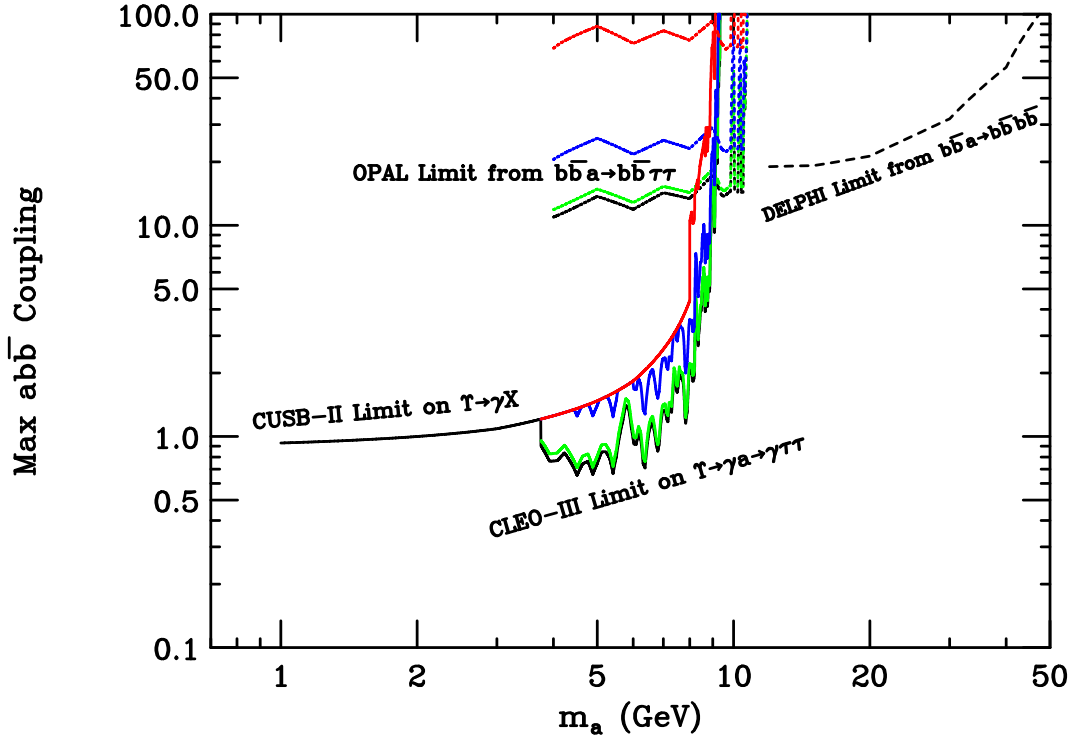}
\end{center}
\caption{Upper limit, $\cmax$, 
 on $|\cabb|$ as a function of $\ma$ coming directly from experimental
 data.
In the case of limits based on $a\to\tauptaum$, curves for
$\rbt=0.5$ (red), 1 (blue), 2 (green), 10 (black) are shown with $\rbt=10$ giving the lowest curves and
$\rbt=0.5$ giving the highest curves.}
\label{coupmaxall}
\end{figure}

Limits on $|\cabb|$ come from $\Upsilon\to \gam a$ decays at $B$
factories and $\epem\to b\anti b a$ production at LEP.  For
$\ma<2\mtau$ the strongest limits come from the old (90\% CL -- all
other limits employed here are 95\% CL) CUSB-II limits~\cite{cusbii}
on $\br(\Upsilon\to \gam X)$, where $X$ is assumed to be visible. For
$2\mtau<\ma<9.2\gev$, the recent CLEO-III~\cite{cleoiii} limits on
$\Upsilon\to \gam a\to \gam \tau\tau$ are the strongest (in
interpreting these limits one must account for the value of $\br(a\to
\tauptaum)$ --- this in turn depends on $\rbt$, but very weakly for
$\rbt\geq 2$, see below).  For $9.2\gev<\ma<\mups$, mixing of the $a$
with various $\eta_b$ and $\chi_0$ bound states becomes
crucial~\cite{Drees:1989du}.  Ref.~\cite{cleoiii} gives results for
$\cmax$ in this $\ma$ range without taking this mixing into account but
notes that their limits cannot be relied upon for $\ma>9.2\gev$.
Whether additional limits can be extracted from lepton
non-universality studies in the $9.2<\ma<\mups$ region is being
studied~\cite{SanchisLozano:2002pm}.  OPAL
limits~\cite{Abbiendi:2001kp} (which assume $\br(a\to
\tauptaum)=1$) on $\epem\to b\anti b\tau\tau$ become
numerically relevant for roughly $9\gev<\ma<2m_b$.
Ref.~\cite{Abbiendi:2001kp} converts these limits to limits on the $a
b\anti b $ coupling using the modeling of \cite{Drees:1989du}.   These
are the only limits in the $\mups<\ma<2\mb$ range and continue to be
relevant up to $12\gev$. Above $\ma=2m_b$ these $ab\anti b$ coupling
limits become quite weak due to the $\eta_b-a$ mixing and the decrease
of $\br(a\to \tau\tau)$. For $\ma\geq 12\gev$, limits on the $ab\anti
b$ coupling can be extracted from $\epem\to b\anti b a\to b\anti b
b\anti b$~\cite{delphi}. The maximum value of $|\cabb|$ allowed by
all these various limits, $\cmax$, is shown in Fig.~\ref{coupmaxall} as a
function of $\ma$ for several values of $\rbt$ ($\rbt=0.5,1,2,10$).  Note
that there is almost no dependence of $\cmax$ on $\rbt$ for $\rbt\geq
2$.  Values of $|\cabb|$ above 50 raise issues of non-perturbativity
of the $a b\anti b$ coupling and are likely to be in conflict with
Tevatron limits on $b\anti b a$ production~\cite{icheptalk}.  $\cmax$
depends on $\rbt$ when the CLEO-III $\Upsilon\to \gamma a\to\gamma
\tauptaum$ or OPAL $b\anti b a\to b\anti b \tauptaum$ limits are the
most relevant.  What is new in this paper is the systematic
incorporation of the $\rbt$ dependence of $\cmax$ and the systematic
incorporation of the $\cmax$ limits in the context of predictions for
$\delta a_\mu$ in a wide class of models.  

In the case of the simple 2HDM(II), where $\cabb=\rbt=\tanb$, values
of $\ma$ for which $\tanb>\cmax(\tanb)$ are not allowed in the model context.
These disallowed regions typically emerge in the range
$\ma<8\gev$ for $\tanb=1$ rising to $\ma\lsim 10\gev$ for higher
$\tanb$; at higher $\tanb$ values they have a complicated structure
that we will discuss later. In addition, a disallowed
region also arises over a limited $\ma$ range starting from $\ma>12\gev$ when
$\tanb\gsim 18$, the larger the value of $\tanb$ the larger the interval.
For example, for $\tanb=50$ the DELPHI limits
imply that the 2HDM(II) is not consistent for $12\lsim \ma\lsim
37\gev$ and the OPAL and Upsilon limits imply that the 2HDM(II) is not
consistent for $\ma<10\gev$. In contrast, for $\tanb=10$ the 2HDM(II)
model is always consistent with the DELPHI limits and is only
inconsistent (with CLEO-III and CUSB limits) for $\ma\lsim 9\gev$.
These 2HDM(II) results are an update of the results obtained in
\cite{Krawczyk:2002df}. The results in all other models, in particular
in the NMSSM context are new.~\footnote{Several months after arXiv
  submission of this paper, similar results for the NMSSM were
  obtained in \cite{Domingo:2008rr}.}

\begin{figure}
\begin{center}
\includegraphics[width=0.65\textwidth,angle=90]{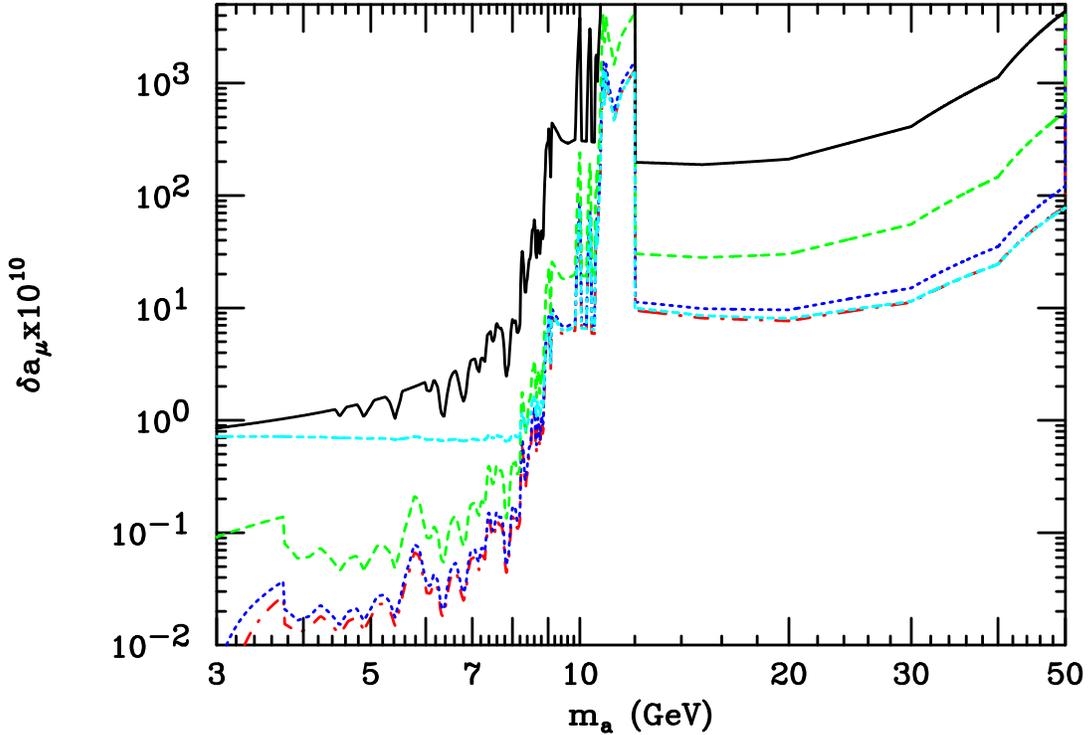}
\end{center}
\caption{The value of $\delta a_\mu$ from CP-odd $a$ loops is plotted as
 a function of $\ma$ for $\rbt=1$ (black, solid), $\rbt=3$ (green,
 dashes), $\rbt=10$ (blue, dots) and $\rbt=50$ (red, long dash,
 short dot, lowest curve),
assuming maximal $ab\anti b$ coupling, $\cmax$, from
Fig.~\protect\ref{coupmaxall}. Also shown as the dotdash cyan curve is the 2HDM(II) prediction
for $\tanb=\cmax(\tanb)$, i.e. the largest possible (self-consistent) choice of $\tanb$ within
the 2HDM(II) context. For $\ma\gsim 9\gev$ this latter curve coincides with the
$\rbt=50$ curve.}
\label{damuvsmauc}
\end{figure}
We will now explore the implications for $a_\mu$.  Since the two-loop
contributions include that with a $t$-loop as well as those with $b$
and $\tau$ loops, we must specify the value of $\catt$ relative to
$\cabb$ in order to compute the contribution of $a$ to $a_\mu$ for a
given $\cabb$ value. In a 2HDM of type-II, including the MSSM and
NMSSM, $\catt=\cotb$ and after including the two-loop diagrams $\delta
a_\mu>0$ for $\ma>2.6,2,0\gev$ if $\tanb>5,3,1$.  In a type-I 2HDM,
$\catt=-\cabb=\cotb$.  Then, the (dominant) top-loop
Barr-Zee type diagram gives a negative contribution to $a_\mu$ and
$\delta a_\mu$ is negative for all $\ma$. Only models
with positive $\rbt^2$ are of relevance for
explaining the observed positive $\Delta a_\mu$. Results for $\delta
a_\mu$ employing the $\cabb=\cmax$ limits as a function of $\ma$ and
$\rbt$ and taking $\rbt=1$, $3$, $10$ and $50$
are plotted in Fig.~\ref{damuvsmauc}. (For $\rbt<1$, simply multiply
the $\rbt=1$ curve by $1/\rbt^2$.)

To a good approximation, $\rbt\geq 50$ is equivalent to dropping the
two-loop diagram containing the top quark and gives the smallest
result.  Since (for positive $\cabb/\catt$) the two-loop top diagram
enters with the same (positive) sign as the $b$ and $\tau$ two-loop
diagrams, the largest $\delta a_\mu$ values are obtained for the
smallest $\rbt$ when using upper limits on the $ab\anti b$ coupling as
input. As a result, we see in Fig.~\ref{damuvsmauc} that for lower
$\rbt$ values ($1<\rbt\lsim 3$) any value of $\ma\gsim 9\gev$ would
make it possible to obtain $\delta a_\mu=\Delta a_\mu\sim 27.5\times
10^{-10}$ for some choice of $\cabb\leq \cmax$.  For $\rbt<0.2$, for
which $\catt$ enters non-perturbative territory, $\delta a_\mu>\Delta
a_\mu$ if $\cabb=\cmax$ for all $\ma$ so that agreement could always
be obtained for some $\cabb<\cmax$.  However, for $\rbt\gsim 10$ the
full discrepancy can only be explained if $10\gev < \ma < 12\gev$ or
$\ma\gsim 36\gev$.  Recall, however, that the value of $\cmax$
extracted from the data in the former region relies on the modeling
for the $a-\eta_b$ mixing employed in the experimental analysis.
Also, for $\ma>36\gev$ and $\rbt\geq 10$, $\delta a_\mu=\Delta a_\mu$
requires non-perturbative $\cabb>50$ values.

\begin{figure}
\begin{center}
\includegraphics[width=0.65\textwidth,angle=90]{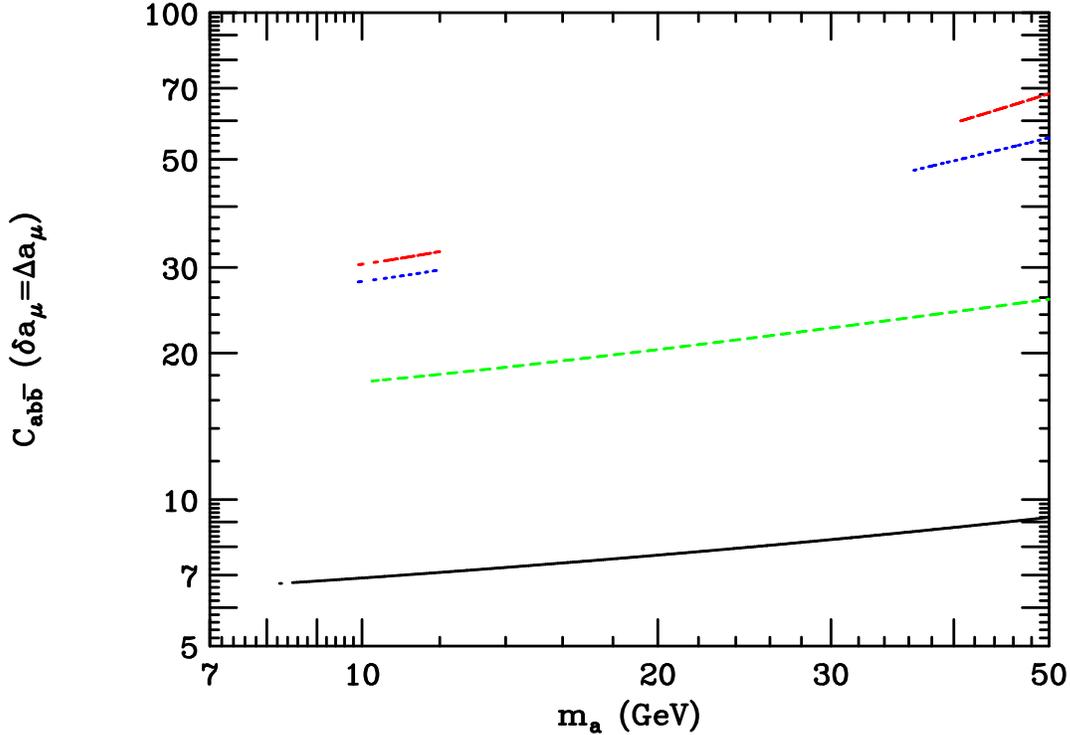}
\end{center}
\caption{The value of of $\cabb$ required in order that $\delta
  a_\mu=27.5\times 10^{-10}$  is plotted as
 a function of $\ma$ for $\rbt=1$ (black, solid), $\rbt=3$ (green,
 dashes), $\rbt=10$ (blue, dots) and $\rbt=50$ (red, long dash, short dot, highest curve),
for those choices of $\ma$ such that the required $\cabb$ is less than
$\cmax$ as plotted in
Fig.~\protect\ref{coupmaxall}. Gaps for any given $\rbt$ curve
correspond to $\ma$ values for which $\cabb>\cmax$ would be required.}
\label{cabbforgminus2vsmauc}
\end{figure}

Of course, it is interesting to know what value of $\cabb<\cmax$ is
needed in order to match the observed $\Delta a_\mu=27.5\times
10^{-10}$ for those $\ma$ and $\rbt$ values for which this is
possible. The results for the general case in which $\cabb$ is not
correlated with $\rbt$ are plotted in Fig.~\ref{cabbforgminus2vsmauc}.
In general, for low values of $\rbt$ (for which the top loop is
a major contributor to $\delta a_\mu$) rather modest values of $\cabb$
will reproduce the observed $\Delta a_\mu$. As $\rbt$ increases, the
bottom loop diagrams must reproduce $\Delta a_\mu$ on their own and
increasingly large values of $\cabb$ are required.
As we shall see, one particularly interesting range of $\ma$ for
$\rbt\geq 10$ is $9.9\gev\lsim \ma\lsim 12\gev$. In
Fig.~\ref{cabbforgminus2vsmauc}, we see that in this $\ma$ range the observed
$\Delta a_\mu=27.5\times 10^{-10}$ is matched for $\cabb$ in the range
$28 \leq \cabb\leq 32$ for $\rbt\geq 10$ when $9.9\gev \lsim \ma \lsim
12\gev$.

\begin{figure}
\begin{center}
\includegraphics[width=0.59\textwidth,angle=90]{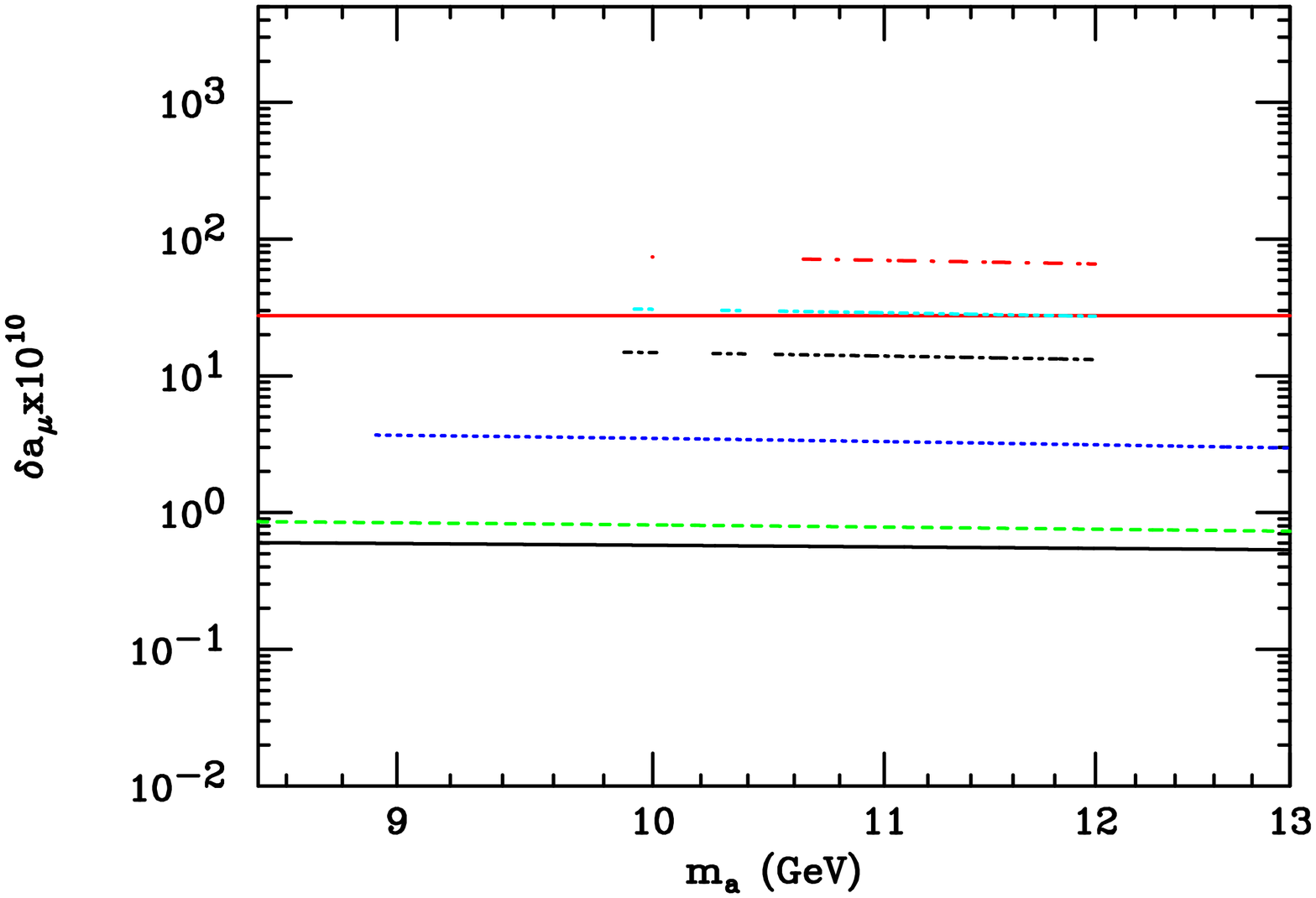}
\includegraphics[width=0.59\textwidth,angle=90]{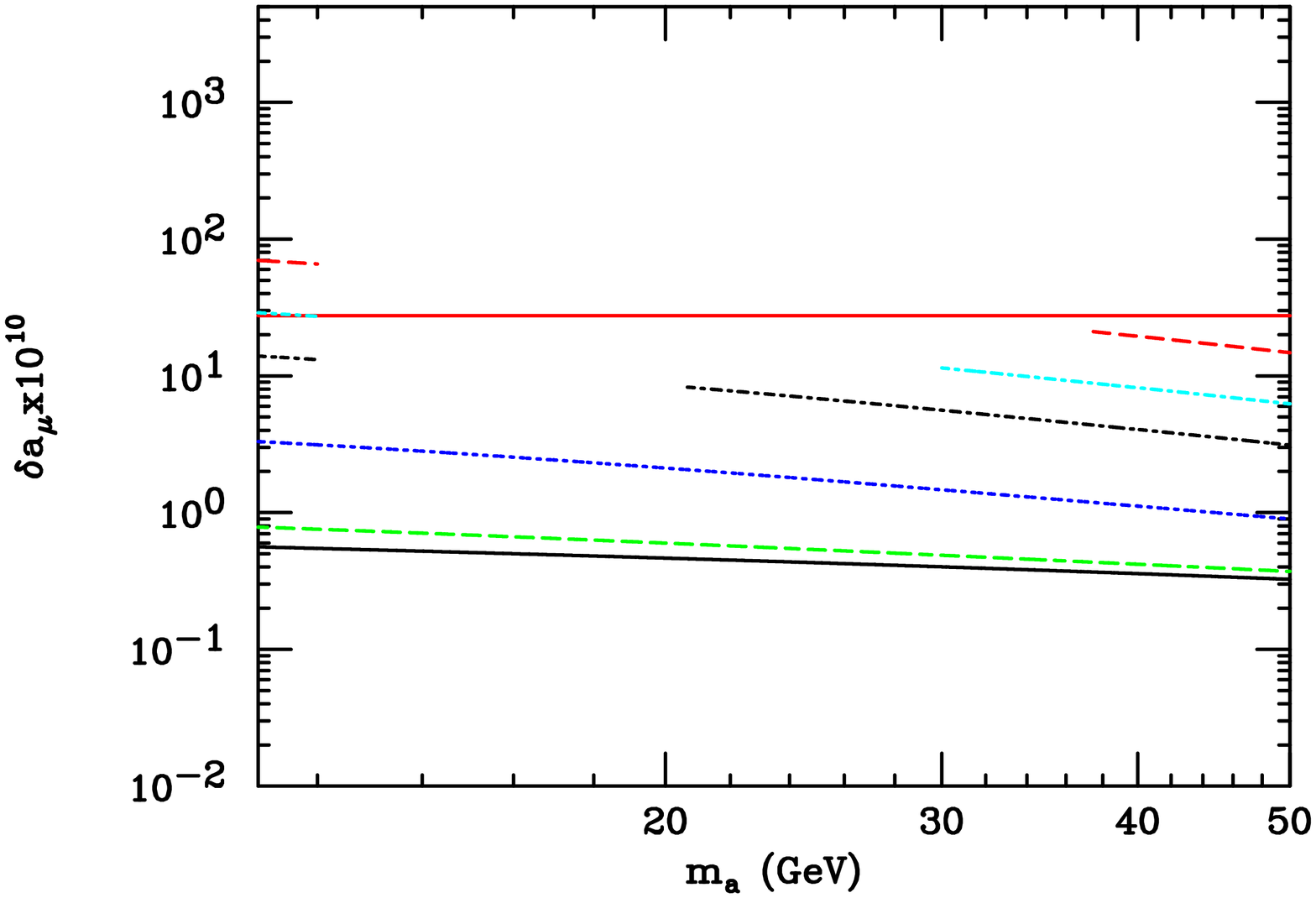}
\end{center}
\vspace*{-.3in}
\caption{The value of $\delta a_\mu$ from CP-odd $a$ loops is plotted as
 a function of $\ma$ for $\tanb=1$ (black, solid), $\tanb=3$ (green,
 dashes), $\tanb=10$ (blue, dots), $\tanb=22$ (black, dotdash), $\tanb=32$ (cyan, dotdash)
 and $\tanb=50$ (red, long dash, short dot, highest curve),
assuming the 2HDM(II) model with $\rbt=\tanb$ and requiring that
$\tanb\leq \cmax(\tanb)$. Omitted regions are those for which $\tanb>\cmax(\tanb)$
as plotted in Fig.~\protect\ref{coupmaxall}. Note the multiple gaps
for the $\tanb=22,32,50$ cases in the $10\gev \leq\ma\leq 11\gev$
region. An intersection of the solid red line
at $\delta a_\mu=27.5\times 10^{-10}$ with a 2HDM(II) curve 
essentially only occurs in the $\tanb=32$ case.
}
\vspace*{-.3in}
\label{damuvsmauc2hdmII}
\end{figure}

The above results are modified in the context of more restrictive
models. Fig.~\ref{damuvsmauc2hdmII} shows the results for $\delta
a_\mu$ in the type-II 2HDM, in which $\cabb=\rbt=\tanb$, obtained for
various $\tanb$ values.  In the type-II 2HDM, the value of $\delta
a_\mu$ is determined once $\tanb$ and $\ma$ are specified. Unlike the
very general case just considered, for which $\rbt$ is not related to
$\cabb$, in the 2HDM(II) context one cannot have large $\cabb$ without
having large $\rbt$, which then minimizes the very important
(positive) top loop contribution.  Thus, the largest $\delta a_\mu$
values are now obtained with large $\tanb$ values.  The possibilities
are also constrained by the requirement that $\tanb$ cannot exceed
$\cmax(\tanb)$. The gaps in the curves of Fig.~\ref{damuvsmauc2hdmII}
are those regions where $\tanb>\cmax(\tanb)$. The result is that in
order to obtain a value of $\delta a_\mu$ of order $27.5\times
10^{-10}$ that also has $\tanb\leq \cmax(\tanb)$ requires a rather
precisely fixed value of $\tanb\sim 30-32$ and $\ma\sim 9.9-12\gev$
(see the $\tanb=32$ dotdash cyan curve). In the context of the most
general CP-conserving type-II 2HDM, any value in the above small range
is not excluded using combined $Zh$ and $ha$ LEP
data~\cite{Abbiendi:2004gn} so long as $\mh\gsim 60\gev$; and, there
are no limits on $\ma$ if $\mh\gsim 100\gev$. Further, contributions
to the precision electroweak observables $S$ and $T$ are tiny if
$\mhh=\mhp$ when $h$ has SM-like $ZZh$ coupling.  As a further remark,
we note from trends as $\tanb$ increases apparent in
Fig.~\ref{damuvsmauc2hdmII} (lower plot) that for $\tanb$ values above
$50$ (i.e. outside the perturbative limit on this coupling) one will
not be able to have $\tanb<\cmax(\tanb)$ in the $\ma<12\gev$ zone, but
that at some largish value of $\ma$ above about $40\gev$ one {\it will} be
able to achieve a match to $\Delta a_\mu$. This is because the DELPHI
limits on $\cabb$ deteriorate so rapidly as $\ma$ increases above
$40\gev$.

As a further
perspective on the 2HDM(II) results, we plot in Fig.~\ref{damuvsmauc}
the largest possible value of $\delta a_\mu$ within the 2HDM(II) as a
function of $\ma$ (the dotdash cyan curve).  This maximal value is obtained when
$\tanb=\cmax(\tanb)$ (i.e. for the largest self-consistent choice of
$\tanb$ such that $\cabb=\tanb$). Again it is apparent that $\delta
a_\mu$ can match (or exceed) $27.5\times 10^{-10}$ in the range
$9.9\lsim \ma \lsim 12\gev$. And, to repeat, matching in this range is
always achieved for $\tanb\sim 30-32$.

The ability to achieve $\delta a_\mu=\Delta a_\mu$ is much more
constrained in the popular Minimal Supersymmetric Model (MSSM). In the
MSSM, the LEP lower limit on $\ma$ is of order $90-100\gev$, depending
upon $\tanb$ and precise model inputs~\cite{Schael:2006cr}.  For
$\ma>90\gev$, $\delta a_\mu=\Delta a_\mu$ is only achievable for
$\cabb=\tanb$ well above the upper bound of $50$ employed here.  (Of
course, if the MSSM sparticles are light, their contributions could
yield the observed $\Delta a_\mu$~\cite{Ellis:2007fu}.)

The Next-to-Minimal Supersymmetric model (NMSSM) provides more fertile
ground.  The NMSSM is obtained by adding a singlet superfield $\what
S$ to the MSSM Higgs superfields $\what H_u$ and $\what H_d$.
Ref.~\cite{Ellis:1988er} was the first to consider the NMSSM Higgs
sector phenomenology in detail.  The scalar component of $\what S$
contains one CP-even and one CP-odd scalar field. The resulting Higgs
sector thus contains three CP-even Higgs bosons ($h_{1,2,3}$) and two
CP-odd Higgs bosons ($a_{1,2}$), all of which can have a singlet
component. A convenient program for exploring the NMSSM Higgs sector
is NMHDECAY~\cite{Ellwanger:2004xm,Ellwanger:2005dv}. We will not
consider contributions to $a_\mu$ from sparticles as recently studied
in~\cite{Domingo:2007dx,Domingo:2008bb}.

The NMSSM is especially attractive in that it allows for the ``ideal''
Higgs sector described earlier with $\mhi\sim 100\gev$, consistent
with LEP data if
$\mai<2m_b$ and $\br(\hi\to \ai\ai)>0.75$. For $\mai>2m_b$, one must
have $\mhi\gsim 110\gev$ to avoid LEP bounds. (But, for
$110\gev\lsim \mhi\lsim 163\gev$, so long as the
$ZZ\hi$ coupling is SM-like the agreement with precision electroweak
data is still within the 95\% CL limit unless only the ``leptonic''
determination of $\sin^2\theta_\ell^{\rm eff}$ is employed in the
precision electroweak analysis; the latter yields a much higher CL for the
overall fit and requires $\mhi\lsim 105\gev$ at 95\% CL --- see \cite{Chanowitz:2008ix} for details).

The most crucial parameter for the NMSSM analysis is $\cta$ defined by
\beq
\ai=\cta a_{MSSM}+\sta a_S,
\eeq
where $a_{MSSM}$ is the CP-odd (doublet) scalar in the MSSM sector of
the NMSSM and $a_S$ is the additional CP-odd singlet scalar of the
NMSSM.  In terms of $\cta$, $\camumu=\cabb=\cta\tanb$ and
$\catt=\cta\cotb$.

Before proceeding, we consider possible constraints from precision
electroweak data. Since the light SM-like $\hi$ already gives good
agreement, the rest of the Higgs sector should give a small
contribution to $S$ and $T$ (assuming sparticle contributions are not
substantial). One finds that if $\mai$ is in the range considered and
$\hi$ is SM-like, then it is typically the case that either $\hii$ or
$\hiii$ is mainly singlet, denoted $h_S$, and the other, denoted here
as $h_D$, is mainly doublet. Further, the $Zh_S\ai$ coupling is very
tiny while the $Zh_D\ai$ coupling is maximal and $\mhp\sim\maii\sim
m_{h_D}$. With these inputs, one finds that the extra contributions
from the Higgs sector to $S$ and $T$ are very small and the excellent
agreement with precision electroweak constraints coming from the $\hi$
is preserved.

\begin{figure}
\begin{center}
\includegraphics[width=0.65\textwidth,angle=90]{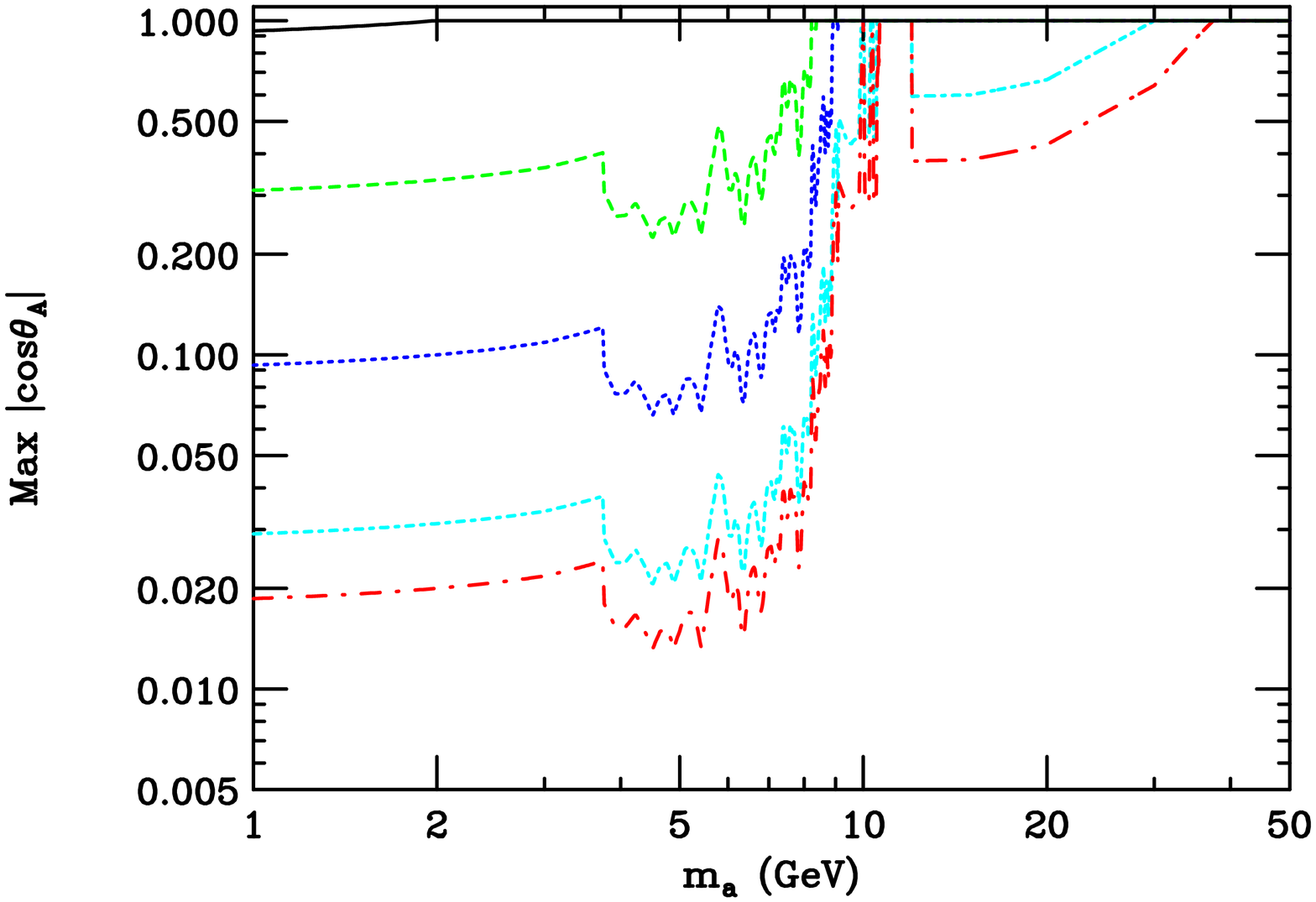}
\end{center}
\caption{$\ctamax$ in the NMSSM (where $\cabb=\cta\tanb$) as a
  function of $\ma$ after requiring $\cta^{max}\tanb=\cmax$ using the
  $\cmax$ values 
  of Fig.~\protect\ref{coupmaxall}.  
 The different curves correspond to $\tanb=1$ (black, solid, upper
 curve), $3$ (green, dashes),
 $10$ (blue, dots), $32$ (cyan, dotdash)  and $50$ (red, long dash, short dot,
 lowest curve).
}
\label{ctamaxvsma}
\end{figure}

Let us now consider $\ai\equiv a$ contributions to $a_\mu$ for various
fixed $\tanb$ values. Then, $\cta$ is constrained by the requirement
that $\cabb=\cta\tanb\leq \cmax$, which constrains $\cta$ to very
small values for low $\ma$ and large $\tanb$.  However, no matter what
the value of $\tanb$, the extra freedom of adjusting $\cta$ does allow
us to avoid gaps in $\ma$ for which $\cabb>\cmax$. This, in turn, will
give us more possibilities for $\delta a_\mu$. Inputting the values of
$\cmax$ as a function of $\ma$ we obtain the results of
Fig.~\ref{ctamaxvsma} for the maximum allowed value of $\cta$ as a
function of $\ma$ for various $\tanb$ values.

\begin{figure}
\begin{center}
\includegraphics[width=0.65\textwidth,angle=90]{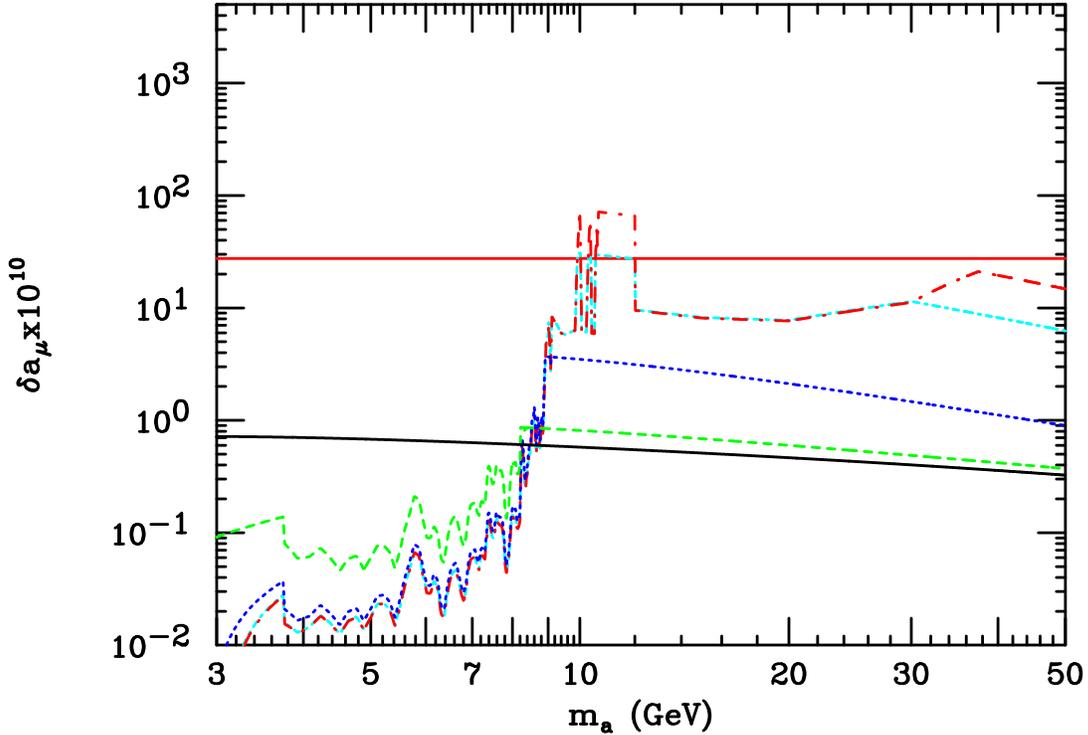}
\end{center}
\caption{The maximum contribution of the CP-odd $a$ to $a_\mu$ as 
  a function of $\ma$ in the NMSSM, for which $\cabb=\cta\tanb$, after
  employing $\cta=\ctamax$ where $\ctamax$ is plotted for different
  $\tanb$ values in Fig.~\protect\ref{ctamaxvsma}.  Curve notation is
  as in Fig.~\protect\ref{ctamaxvsma}. The horizontal solid red line
  is located at $\delta a_\mu =27.5\times 10^{-10}$.}
\label{damuvsma}
\end{figure}

We now turn to the resulting NMSSM predictions for $a_\mu$.  The value of $\delta
a_\mu$ is largest for $\cta=\ctamax$.  The resulting values of $\delta
a_\mu$ are plotted as a function of $\ma$ in Fig.~\ref{damuvsma}.
As in the generic case, the strong constraints from Upsilon physics
imply that significant contributions to $a_\mu$ are not possible until
$\ma$ exceeds roughly $9.2\gev$.  To understand why $\delta a_\mu$
increases with increasing $\tanb$ for $\ma>12\gev$, whereas it
decreases with increasing $\tanb$ for low $\ma$, we first note that
the 2-loop, top-loop contribution to $\delta a_\mu$ is independent of
$\tanb$ (because of a $\camumu\catt$ structure that is
$\tanb$-independent), whereas the 2-loop bottom-loop contribution
increases as $\tan^2\beta$ (because of a $\camumu\cabb\propto\tan^2\beta$
structure). Numerically, before including the extra $\tan^2\beta$
factor for the latter, the 2-loop, top-loop contribution is much
larger than the 2-loop, bottom-loop contribution. Of course, both
contributions are multiplied by $(\cta)^2$. Thus, when $\cmax$ is
independent of $\tanb$ and $\ctamax=1$ (as for $\ma>12\gev$ and
$\tanb\lsim 20$) the resulting $\delta a_\mu$ will always increase
with $\tanb$. However, at low $\ma$, the very strong Upsilon
constraints on $\cabb$ imply that $\ctamax$ rapidly decreases with
increasing $\tanb$ which suppresses the numerically more important
2-loop, top-loop contribution resulting in smaller $\delta a_\mu$ as
$\tanb$ increases.

From Fig.~\ref{damuvsma}, we observe
that the maximal $\delta a_\mu$ can exceed $\Delta a_\mu=27.5\times
10^{-10}$ for $9.9\gev\lsim \ma\lsim 12\gev$ if $\tanb\geq 32$, with
an almost precise match to this value of $\Delta a_\mu$ for $\tanb=32$
(or for $\tanb$ as low as $\tanb=30$ --- see the 2HDM discussion).
For $\tanb=50$, one can match $\Delta a_\mu$ by using a value of
$\cta$ below $\ctamax$. (As discussed below, the fact that matching is
possible for $9.9\gev\lsim \ma\lsim 2m_B$ is particularly interesting in the
context of the ideal Higgs scenario.) Further, the maximal $\delta
a_\mu$ is in the $7-20\times 10^{-10}$ range for $12\gev < \ma\lsim
48\gev$ for $\tanb=32$ and for $12\gev < \ma\lsim 70\gev$ for
$\tanb=50$.

At this point, it is worth stressing the other desirable features of
the $\mh\sim 100\gev$, $\ma\lsim 2m_B$, $\br(h\to aa)>0.75$ scenario
as discussed
in~\cite{Dermisek:2005ar,Dermisek:2005gg,Dermisek:2006wr,Dermisek:2007yt}.
These references examined the degree to which obtaining the observed
value of $m_Z$ requires very precisely tuned values of the GUT scale
parameters of the MSSM and NMSSM.  One finds that in any
supersymmetric model this finetuning is always minimized for GUT scale
parameters that yield a SM-like $h$ with $\mh\leq 100\gev$, something
that is only consistent with LEP data if the $h$ has unexpected decays
that reduce the $h\to b\anti b$ branching ratio while not contributing
to $h\to b\anti b b\anti b$ (also strongly constrained by LEP data).
A Higgs sector with a light $a$ for which $\br(h\to aa)>0.75$ and with
$\ma$ small enough that $a$
decays to $B\anti B$ final states are disallowed (i.e. $\ma <
10.56\gev$) provides a very natural possibility for allowing minimal
finetuning. The NMSSM provides one possible example.

In conclusion, the combined limits from $\Upsilon$ decays and $b\anti
ba$ Yukawa production at LEP, along with the perturbativity
requirement of $\cabb<50$, imply that the entire $a_\mu$ discrepancy
of $\Delta a_\mu\sim 30\times 10^{-10}$ cannot have a purely Higgs
sector explanation without going beyond the MSSM. In the
less-constrained NMSSM, achieving $\delta a_\mu\sim \Delta a_\mu$
requires relatively high $\tanb$ and a value of $\ma$ between about
$10\gev$ and $2m_B$.  On the one hand, this is a highly motivated
$\ma$ region in the NMSSM since, as described earlier, it would allow
an ``ideal'' SM-like $h$ with $\mh\lsim 100\gev$ decaying mainly via
$h\to aa\to 4 \tau$. Such an $h$ would escape LEP limits while
allowing for low $m_Z$-finetuning. However, on the other hand, in the
NMSSM $\ma<2m_B$ most naturally arises when close to the $U(1)_R$
symmetry limit.  In this case, the $a$ is mainly singlet, implying
that $\cta$ is small and that $\cabb=\cta\tanb$ is typically
$\calo(1)$~\cite{Dermisek:2006wr}, whereas $\cabb\sim 30$ is needed to
match the observed $\Delta a_\mu$.

Nonetheless, the possibility that a CP-odd $a$ with $10\gev\lsim
\ma\lsim 12\gev$ could explain the $a_\mu$ anomaly should be taken seriously,
Thus, finding techniques to experimentally probe for an $a$ in the
$10\gev<\ma<12\gev$ region should be a high priority.  Such new
techniques could either end up limiting $\cabb$ sufficiently that
$\Delta a_\mu$ cannot be explained in the 2HDM(II) or NMSSM frameworks
or else actually allow a discovery of  a light $a$. Of course, this is a region in which
$\eta_b-a$ mixing will surely be a complication.

As an aside, one must not forget that in supersymmetric models sparticle loops could
have two important roles: (i) they could directly yield large
contributions to $a_\mu$; and (ii) they could modify the relations
between $\camumu$, $\catautau$, $\cabb$ and $\catt$.

If one goes beyond the MSSM and NMSSM Higgs sectors to the more
general type-II 2HDM, then, keeping $\cabb<50$, only an $a$ with
$10\gev<\ma<12\gev$ with $\cabb\sim 30-32$ could give $\delta
a_\mu=\Delta a_\mu$. (A type-I 2HDM gives negative $\delta a_\mu$ that
is large for $\ma>8\gev$ if $\cabb=\cmax$ and is therefore strongly
disfavored by the observed positive $\Delta a_\mu$.)  

Obtaining the observed $\Delta a_\mu$ in the most general Higgs model
for which the $a b\anti b$ coupling magnitude is disconnected from the
ratio $\rbt^2$ of the $ab\anti b$ to $at\anti t$ couplings is
generically possible so long as $\rbt^2>0$. For $\rbt= 1$, $\ma>8\gev$
and a relatively modest value of $\cabb$ (well below the maximum
allowed) will yield $\delta a_\mu=\Delta a_\mu$. As $\rbt$ increases,
the required $\cabb$ increases.  For larger $\rbt$, there are regions
of $\ma$ for which the required $\cabb$ exceeds the upper experimental
bound, $\cmax$. Further, $\delta a_\mu=\Delta a_\mu$ cannot be
achieved above an $\rbt$-dependent maximum $\ma$ if $\cabb<50$ is
imposed.  For $\rbt<0.2$, even very low values of $\ma$ will yield the
observed $\Delta a_\mu$ for an appropriate choice of $\cabb<\cmax$.

\acknowledgments 

This work was supported by U.S. DOE grant No. DE-FG03-91ER40674. This
research was supported in part by the National Science Foundation
under Grant No. PHY05-51164 while at KITP and by the Aspen Center for
Physics. JFG thanks R. Dermisek, T. Han, B. McElrath, and P. Osland
for their comments on the manuscript.

\end{document}